\documentstyle[12pt]{article}
\def\beq{\begin{equation}}
\def\eeq{\end{equation}}

\begin{document}

\renewcommand{\thefootnote}{\fnsymbol{footnote}}
\newcommand{\ket}{\rangle}
\newcommand{\bra}{\langle}

\begin{center}
{\LARGE {\sc Essay Review}}\\[7mm]

{\Large{\bf Interpreting the Quantum World}}\\[7mm]

{\Large{\sl Asher Peres}\footnote{Department of Physics,
Technion---Israel Institute of Technology, 32\,000 Haifa,
Israel}}\end{center}

\bigskip\noindent Jeffrey Bub, {\it Interpreting the Quantum World\/}
(Cambridge: Cambridge University Press, 1997), xiv + 298~pp. ISBN
0-521-56082-9 hardback \pounds35.00 (US\$49.95)\bigskip

\renewcommand{\thefootnote}{\arabic{footnote}}
\setcounter{footnote}{0}

\noindent The object of this book is the physical interpretation of the
abstract formalism of quantum theory. This issue has been controversial
from the early days of quantum mechanics, more than 70 years ago. Many
of the best minds struggled with this problem, only to reach conflicting
conclusions. Obviously, there is no similar interpretation problem for
classical mechanics, because the mathematical symbols that appear in the
latter simply coincide with experimentally observable quantities. On the
other hand, the quantum formalism is based on a complex vector space in
which the dynamical evolution is generated by unitary operators.
Everyone agrees on how to manipulate the mathematical symbols; the
thorny problem is to relate them to the observable physical reality.

The traditional answer is to introduce `observers' who sense the quantum
world by interacting with it. While they are engaged in that
interaction, the observers must obey quantum dynamics---this is needed
for consistency of the formalism. Yet, after completion of the measuring
process, the same observers must be given a mundane, classical,
objective description, so that the `quantum measurement' ends with a
definite result, as we experience in everyday's life. Quantum mechanics
itself does not predict, in general, that result. It predicts only {\it
probabilities\/} for the various possible outcomes of a measurement,
once we specify the procedure used for the preparation of the physical
system. This {\it ad hoc\/} approach is sufficient for the purposes of
experimental physics, and it can even be rationalized by some
theoretical physicists (including the author of this review). However,
it is considered as unacceptable by philosophers of science. Bub's book
gave me an opportunity to understand why.

Bub's goal is to liberate the quantum world from its dependence on
observers.  Various possibilities are carefully examined. The book
contains an amazing wealth of information, including numerous excerpts
of correspondence between Einstein, Schr\"odinger, Pauli, Born, and
others. I have particularly been impressed by the two long chapters (75
pages) which analyze in exhaustive detail the celebrated `no go'
theorems of Bell and of Kochen and Specker, namely the contradictions
that would arise in any attempt to supplement the quantum wave function
by additional `hidden' variables, whose objective values would
unambiguously determine the result of any quantum measurement (the wave
function itself supplies only {\it statistical\/} information on the
possible outcomes of physical experiments). These two chapters are a
welcome update to Redhead's (1987) treatise on this subject. The other
chapters of the book analyze various interpretations of quantum theory,
on which I shall say more later.

On the negative side, I must confess that I have often been irritated by
what may be, after all, only a matter of terminology. Bub repeatedly
uses the expression `the value of a dynamical variable.' While reading
the first pages of the book, I tried to count how many times this
expression appears, but I quickly lost the count. In classical
mechanics, a dynamical variable indeed has a definite value at each
point of phase space. Specifying a point in phase space is the standard
way of indicating the state of a physical system. However, in quantum
mechanics, a dynamical variable is represented by a Hermitian matrix
(or, more generally, by a self-adjoint operator). It is manifestly
pointless to attribute to it a numerical value. The founding fathers
(Einstein, Schr\"odinger, and others), who had been bred in classical
concepts, could be excused for trying to force the classical language on
quantum theory. There is no excuse for that today.

The real issue that I can see at this point is whether Bub's
interpretation problem is artificial (namely, it results from imposing a
classical language for the description of quantum phenomena), or whether
there is a genuine fundamental difficulty rooted in inadequate physical
concepts. At the very beginning of his book (page~2), Bub defines the
result of the measuring process as follows:

\begin{quote}{\small In the orthodox interpretation, neither the
measured observable nor the pointer reading have determinate values,
after a suitable interaction that correlates pointer readings with
values of the measured observable. This is the measurement problem of
quantum mechanics.}\end{quote}

\noindent In other words, quantum mechanics does not provide a
satisfactory description of the measuring process.

The entire book is devoted to a discussion of various attempts to solve
the measurement problem. The tacit assumption made by Bub (as well as by
many other authors who tried to come to grips with that problem) is that
the wave function is a genuine physical entity, not just an intellectual
tool invented for the purpose of computing probabilities. Even if the
wave function is not an ordinary physical object, it still has
ontological meaning: it represents the factual physical situation, not
only our subjective knowledge of nature. 

Basically, the problem is whether quantum mechanics is a theory of
physical reality, or one of our perception of the physical world. For
the advocates of the realistic alternative, who wish to give a
consistent dynamical description to the measuring process, the source of
the difficulty is clear: the evolution that we know to write for the
quantum mechanical wave function during a measurement does not
correspond with what is actually seen happening in the real world.

In the theoretical laboratory, wave functions are routinely employed by
physicists as mathematical tools, which are useful for predicting
probabilities for the various possible outcomes of a measuring process.
The natural question is: can a more detailed description of that process
be given? In order to investigate this problem, the nature of a quantum
measurement must be clearly understood. The outcome of a measurement is
more than the mere occurrence of an unpredictable event, such as the
blackening of a grain in a photographic plate, or an electric discharge
in a particle detector. To be meaningful, these macroscopic events must
be accompanied by a {\it theoretical interpretation\/}, and the latter
is always at least partly formulated in a {\it classical\/} language.
For example, the Stern-Gerlach experiment is interpreted as the
measurement of a magnetic moment, because it could indeed be such a
measurement if we sent little compass needles through the Stern-Gerlach
inhomogeneous magnet, instead of sending silver atoms. When nuclear
physicists measure cross sections, they assume that the nuclear fragment
trajectories are classical straight lines between the target and the
various detectors. Without this assumption, the macroscopic positions of
the detectors could not be converted into angles for the differential
nuclear cross sections. Quantum theory appears only at the next stage,
to explain, or predict, the possible values of the magnetic moment, the
cross sections, the wavelengths, etc.

The measuring process is concluded by establishing an objective
indelible {\it record\/}. The record must be objective (i.e., all
observers shall agree about its contents), even if the `physical
quantity' to which that record claims to refer is not. Moreover, to have
a meaningful result, we must {\it interpret\/} the experimental outcomes
produced by our equipment. As I just mentioned, this is done by
constructing a theoretical model whereby the behavior of the macroscopic
equipment is described by a few degrees of freedom, interacting with
those of the microscopic system under observation. This is the procedure
that is called a `measurement' of the microscopic system. The logical
conclusion was drawn long ago by Kemble (1937):

\begin{quote} {\small We have no satisfactory reason for ascribing
objective existence to physical quantities as distinguished from the
numbers obtained when we make the measurements which we correlate with
them. There is no real reason for supposing that a particle has at every
moment a definite, but unknown, position which may be revealed by a
measurement of the right kind, or a definite momentum which can be
revealed by a different measurement. On the contrary, we get into a maze
of contradictions as soon as we inject into quantum mechanics such
concepts carried over from the language and philosophy of our
ancestors \ldots\ It would be more exact if we spoke of `making
measurements' of this, that, or the other type instead of saying that
we measure this, that, or the other `physical quantity.'} \end{quote}

However, the above statements are not the interpretation that Bub takes
as the basis for his book. Rather, he considers the orthodox
interpretation of quantum mechanics as the one formulated by Dirac and
von Neumann, with `quantum jumps' (a.k.a.\ collapses of the wave
function). According to that interpretative principle, observables come
to have determinate values when they are actually measured. Bub
thoroughly criticizes that orthodoxy, and rightfully so. He writes
(page~34):

\begin{quote}{\small The upshot of this analysis is that `measurement,'
in the sense required by the orthodox interpretation of the dynamical
state as yielding probabilities of measurement outcomes, can't be
understood dynamically. It follows that an observable that has no
determinate value cannot come to have a determinate value as the result
of a measurement understood as a dynamical interaction between a
measured system and a measuring instrument, and so the requirement of
the orthodox interpretation that observables come to have determinate
values when measured has no dynamical justification.}\end{quote}

\noindent As an example,

\begin{quote}{\small [Schr\"odinger's] cat will be neither alive nor
dead after the measurement interaction, according to the orthodox
interpretation.}\end{quote}

\noindent Since the `orthodox' interpretation is so badly flawed, what
can be acceptable alternatives? The introduction of additional, hidden
variables has been proposed, but the latter lead to new problems that
are the subject of the next chapters.

Chapter 2 (Bell's `no go' theorem) starts with a lengthy discussion of
the Einstein-Podolsky-Rosen (1935) incompleteness argument. Bub then
gives not one, but several proofs of Bell's theorem, under slightly
different physical assumptions. All these proofs are thoroughly analyzed
from the logical point of view. Unfortunately, there is no mention of
possible loopholes in the corresponding experimental tests (these tests
are an unalienable part of the Quantum World). Loopholes in the tests
were pointed out by Santos (1991, 1992) and by many others, and they
also lead to exquisite logical problems.

Moreover, Bohr's (1935) rebuttal of the article of Einstein, Podolsky,
and Rosen is almost completely ignored. It is not mentioned at all in
Chapter 2, nor in Chapter 3 (which is about Bell's other theorem,
a.k.a.\ the Kochen-Specker theorem). Bub finally acknowledges Bohr's
analysis of the  Einstein-Podolsky-Rosen article only near the end of
his book (pp.~198--200), and there he writes:

\begin{quote}{\small The heart of Bohr's response to the EPR argument is
in a footnote \ldots\ [and] in a further footnote that comments on the
original footnote. No part of the paper other than the two footnotes and
the brief remarks preceding the second footnote specifically addresses
the EPR argument.}\end{quote}

\noindent I find it rather disappointing that the discussion of Bohr's
article is restricted to these two strictly technical footnotes. I have
always considered that article as one of the landmarks of quantum
theory!

Chapter 3 (The Kochen and Specker `no go' theorem) contains the most
comprehensive anthology I have seen on this subject. In its original
form, the Kochen-Specker theorem asserts that, in a Hilbert space with
a finite number of dimensions, $d\geq3$, it is possible to construct a
set of $n$ projection operators, which represent yes-no questions about
a quantum system of dimensionality $d$, such that none of the $2^n$
possible yes or no answers is compatible with the sum rules of quantum
mechanics. Namely, if a subset of mutually orthogonal projection
operators sums up to the unit matrix, one and only one of the
corresponding answers ought to be yes. The physical meaning of this
theorem is that there is no way of introducing noncontextual `hidden'
variables which would ascribe definite outcomes to these $n$ yes-no
tests. This conclusion holds irrespective of the preparation (i.e., the
quantum state) of the system being tested.

Bub provides numerous examples, starting with a hitherto unpublished
letter written in 1965 by the logician Kurt Sch\"utte to Ernst Specker.
The latest and most economical proofs are due to Kernaghan (1994) and to
Cabello {\it et al.\/} (1996).

It is flattering that several illustrations from my book (Peres, 1993)
are reproduced here. Yet, I am again disappointed that Bub chose to
refer only to these technical points in my book, and totally ignored the
rest of my views on the physical interpretation of quantum theory. (O
well, I have no right to complain. I wrote in the preface: `This is not
a book on the philosophy of science.')

The problem of interpretation is the subject of Chapter 4, which is the
heart of the book. Bub points out (page 116), and I wholeheartedly
agree, that

\begin{quote}{\small von Neumann was a learned mathematician {\it par
excellence\/}, but he largely abdicates this r\^ole in his discussion of
measurement in quantum mechanics in favour of that of a (rather
uncritical) metaphysician.}\end{quote}

\noindent Indeed, the `orthodox' interpretation that was set up by Dirac
and von Neumann states that (pp.~117, 118)

\begin{quote}{\small an observable has a determinate (definite, sharp)
value for a system in a given quantum state if and only if the state is
an eigenstate of the observable \ldots\ The orthodox interpretation
leads to the measurement problem, which Dirac and von Neumann resolve
formally by invoking quantum jumps or a projection postulate that
characterizes the `collapse' or projection of the quantum state of the
system onto an eigenstate of the measured observable. Dynamical
`collapse' interpretations of quantum mechanics keep the orthodox
interpretation of the quantum state and modify the unitary Schr\"odinger
dynamics of the theory to achieve the required state evolution for both
measurement and nonmeasurement interactions.}\end{quote}

Chapter 4 is relatively short and is mostly devoted to a rigorous proof
of a uniqueness theorem for the `no collapse' interpretations of quantum
mechanics. The theorem states that, subject to certain natural
constraints, all these interpretations can be uniquely characterized and
reduced to the choice of a particular preferred observable as
determinate. The preferred observable and the quantum state of the
system define a non-Boolean `determinate' sublattice in the lattice of
all subspaces of Hilbert space---namely, the sublattice of propositions
that can be true or false. The actual properties of the system are
selected by a two-valued homomorphism (a yes-no map) on the determinate
sublattice, so that the range of possibilities for the system is defined
by the set of two-valued homomorphisms on that sublattice. From this
`modal' perspective, the quantum dynamical state is distinct from the
`property state' defined by the two-valued homomorphism. Different
choices for the preferred determinate observable correspond to different
`no collapse' interpretations of quantum mechanics.

All this is necessary preparatory work for Chapters 5 and 6, both
entitled `Quantum mechanics without observers.' Bub wants to get rid of
these fictitious observers, in spite of the fact that they appear to be
quite harmless. (They are like the ubiquitous observers who send and
receive light signals in textbooks on the theory of relativity.) On the
other hand Bub still endeavours to assign numerical values to
observables, or truth values to lattices of propositions. These are
notions that have been borrowed from the classical world, and I don't
see why quantum reality, whatever it is, has to be described in terms of
observables or lattices of propositions. The fundamental conundrum in
the quantum formalism is not there. It is that quantum mechanics permits
the occurrence  of all possible events (with definite probabilities),
but in our consciousness there is only one world.

Among attempts to solve the measurement problem, Bub examines Bohmian
mechanics, non-ideal measurements, and the environmental `monitoring'
that leads to decoherence. He gives a concise proof of the
tridecompositional theorem (a generalization of the Schmidt
decomposition), which is essential for the various modal interpretations
(page~178):

\begin{quote}{\small All these modal interpretations share \ldots\ the
feature that an observable can have a determinate value even if the
quantum state is not an eigenstate of the observable, so that they
preserve the linear, unitary dynamics for quantum states without
requiring the projection postulate to validate the determinateness of
pointer readings and measured observable values in quantum measurement
processes.}\end{quote}

\noindent On the other hand, Bohmian mechanics, as defined by Bohm and
and Hiley (1993) is a reformulation of quantum mechanics in terms of
`beables':

\begin{quote}{\small This theory is formulated basically in terms of
what Bell has called `beables' rather than of `observables.' These
beables are assumed to have a reality that is independent of being
observed or known in any other way. The observables therefore do not
have a fundamental significance in the theory but rather are treated as
statistical functions of the beables that are involved in what is
currently called a measurement.}\end{quote}

\noindent Likewise for Bell (1987), the `beables' of a physical theory
are equivalent to its `elements of reality' as defined by Einstein,
Podolsky, and Rosen (1935). Bell writes:

\begin{quote}{\small In particular we will exclude the notion of
`observable' in favour of that of `beable.' The beables of the theory
are those elements which might correspond to elements of reality, to
things which exist. Their existence does not depend on `observation.'
Indeed observation and observers must be made out of
beables.}\end{quote}

\noindent Yet, as Bell himself had strikingly shown many years earlier,
these elements of reality engender serious difficulties when we want to
retain the commonly accepted assumptions on the locality of physical
phenomena (see the `no go' theorem in Chapter 2).

Chapter 7 is entitled `Orthodoxy,' but it is {\it not\/} about the
`orthodox' interpretation of Dirac and von Neumann, that was discussed
above. The chapter starts with a section on the so-called Copenhagen
interpretation. [I always refrain from using that terminology, because
the `Copenhagen interpretation' comes in many variants, which may be in
complete opposition to each other. Compare for example the reviews by
Ballentine (1970) and Stapp (1972).]

In the quantum folklore, the Copenhagen interpretation is linked to the
uncertainty relations, which are a special case of the complementarity
principle. Bub points out minor differences between the points of view
of Bohr and Heisenberg. He mentions Wigner's dualism and Wheeler's
observer-participancy, and then he asserts (page 191):

\begin{quote}{\small It is generally recognized---at least, by all but
the most recalcitrant positivists---that the mere fact that measurements
disturb what we measure does not preclude the possibility that
observables have determinate values, or even that measurements might be
exploited to reveal these values in suitably designed measurement
contexts. The `disturbance' terminology itself suggests the existence
of determinate values for observables, prior to measurement, that are
`disturbed' or undergo dynamical change in physical interactions.}
\end{quote}

\noindent For the opinion of some recalcitrant positivists on how to
actually define `disturbance' in quantum mechanics, see Fuchs and Peres
(1996).

Bub does not do justice to Bohr's point of view. He barely mentions
Bohr's insistence that measuring instruments must always be described on
classical lines. This classical-quantum duality, not the particle-wave
duality, is the most significant feature of Bohr's interpretation (Bohr,
1949):

\begin{quote}{\small In a lecture on that occasion [Como, 1927], I
advocated a point of view conveniently termed `complementarity,' suited
to embrace the characteristic features of individuality of quantum
phenomena, and at the same time to clarify the peculiar aspects of the
observational problem in this field of experience. For this purpose, it
is decisive to recognize that, [author's italics] {\it however far the
phenomena transcend the scope of classical physical explanation, the
account of all evidence must be expressed in classical terms\/}. The
argument is simply that by the word `experiment' we refer to a situation
where we can tell others what we have done and what we have learned and
that, therefore, the account of the experimental arrangement and the
results of the observations must be expressed in unambiguous language
with suitable application of the terminology of classical
physics.}\end{quote}  

Bohr was very careful and never claimed that there were in nature two
different types of physical systems. All he said was that we had to use
two different (classical or quantum) {\it languages\/} in order to
describe different parts of the world. The peculiar property of the
quantum measuring process is that we have to use {\it both\/}
descriptions for the {\it same\/} object: namely, the measuring
apparatus obeys quantum dynamics while it interacts with the quantum
system under study, and at a later stage the same apparatus is
considered as a classical object, when it becomes the permanent
depositary of information. This dichotomy is the root of the quantum
measurement dilemma: there can be no unambiguous classical-quantum
dictionary. There can only be an {\it approximate\/} correspondence
between the two languages. That mismatch is the source of the so-called
`quantum uncertainties.' 

Bohr also was sometimes quite elusive. He never explicitly treated a
measurement as an interaction between two quantum systems followed by a
collapse of their wave function, and he could thereby completely elude
the measurement problem. Indeed Bub acknowledges (page 195) that

\begin{quote}{\small Complementarity, too, can be understood as a type
of `no collapse' interpretation. From the perspective of the uniqueness
theorem, Bohr's complementarity and Einstein's realism (in the `beable'
sense, stripped of Einstein stringent separability and locality
requirements) appear as two quite different proposals for selecting the
preferred determinate observable: either fixed, once and for all, as the
realist would require, or settled pragmatically by what we choose to
observe. (So complementarity is not an {\it observer-free\/} `no
collapse' interpretation.)}\end{quote}

Chapter 8 (The new orthodoxy) discusses various mutations of quantum
mechanics that are currently popular: environment-induced decoherence
(which cannot solve the measurement problem), the many-worlds
interpretation, and the method of `consistent histories' which is
related to both of them. If taken seriously, the many-worlds
interpretation leads to bizarre declarations, such as (page 227):

\begin{quote}{\small Suppose that Eve is competent to report her mental
state when it is not an eigenstate of some definite belief about
$S$-spin \ldots\ Eve will respond that she has a definite belief about
the spin of the electron even when she in fact has no definite belief
\ldots\ Eve is apparently going to be radically deceived even about what
her own occurrent mental state is.}\end{quote}

\noindent The subject is no longer physics. It is psychology, or perhaps
psychiatry, well beyond my limited area of competence.

The theory of consistent histories is more technical, which makes it
vulnerable to various technical claims of inconsistency. Its aim is to
encompass the entire universe, including some `quasi-classical domains'
that play a role analogous to that of the Copenhagen observers. Here is
the right place to ask whether there can be a meaningful non-Copenhagen
variant of quantum theory that applies to everything in the universe, in
particular to the atoms in my brain. It seems to me that such a quest
makes no sense and can only lead to self-referential delusions, as in
the above quote. The very idea of a wave function of the universe is
beyond my comprehension.

Returning to physics, a completely different problem is whether we can
consider just a few collective degrees of freedom of the universe, such
as its radius, mean density, total baryon number, etc., and apply
quantum theory to these degrees of freedom, which do not include the
observer and other insignificant details. This seems legitimate: this is
not essentially different from quantizing the magnetic flux and the
electric current in a SQUID, and ignoring the atomic details. You may
object that there is only one universe, but likewise there is only one
SQUID in my laboratory. For sure, I can manipulate that SQUID more
easily than I would manipulate the radius of the universe. Still, that
SQUID is {\it unique\/}. There is no difference in principle.

The last chapter (Coda) summarizes the arguments of the book, and a long
appendix (Some mathematical machinery) comes to help non-initiated
readers. In conclusion, there is much to learn from Bub's book, and I
recommend it without hesitation to all those who are interested in the
foundations of quantum theory.

However, I must add that the contents of this book did not fulfill some
of my expectations. It has an attractive title, {\it Interpreting the
Quantum World\/}, but is this a genuine interpretation? My dictionary
(Webster, 1974) defines `interpret': 1. to explain the meaning of, make
understandable \ldots\ 2. to translate \ldots\ (there also are other
definitions, not relevant here). To interpret quantum theory, to explain
its meaning, to make it understandable, the way has been shown to us: we
have to {\it translate\/} the abstract mathematical formalism in such a
way that `we can tell others what we have done and what we have learned
and [this] must be expressed in unambiguous language with the
terminology of classical physics.'

\begin{center}{\bf References}\end{center}\frenchspacing

\begin{description}

\item Ballentine, L. E. (1970) `The Statistical Interpretation of
Quantum Mechanics' {\it Reviews of Modern Physics\/} {\bf 42}, 358--381.

\item Bell, J. S. (1987) {\it Speakable and Unspeakable in Quantum
Mechanics\/} (Cambridge: Cambridge University Press).

\item Bohm, D. and Hiley, B. J. (1993) {\it The Undivided Universe: An
Ontological Interpretation of Quantum Theory\/} (London: Routledge).

\item Bohr, N. (1935) `Can Quantum-Mechanical Description of Physical
Reality be Considered Complete?' {\it Physical Review\/} {\bf 48},
696--702.

\item Bohr, N. (1949) `Discussion with Einstein on Epistemological
Problems in Atomic Physics' in P.~A.~Schilpp (ed.) {\it Albert Einstein,
Philosopher-Scientist\/} (Evanston: Library of Living Philosophers)
pp.~201--241. 

\item Cabello, A., Estebaranz, J. M., and Garc\'{\i}a-Alcaine, G.,
(1996) `Bell-Kochen-Specker Theorem: a Proof with 18 Vectors' {\it
Physics Letters A} {\bf 212}, 183--187.

\item Einstein, A., Podolsky, B., and Rosen, N. (1935) `Can
Quantum-Mechanical Description of Physical Reality be Considered
Complete?' {\it Physical Review\/} {\bf 47}, 777--780.

\item Fuchs, C. A. and Peres, A. (1996) `Quantum State Disturbance
versus Information Gain: Uncertainty Relations for Quantum Information'
{\it Physical Review A\/} {\bf 53}, 2038--2045.

\item Kemble, E. C. (1937) {\it The Fundamental Principles of Quantum
Mechanics\/} (New York: McGraw-Hill, reprinted by Dover) pp.~243--244.

\item Kernaghan, M., (1994) `Bell-Kochen-Specker Theorem for 20 Vectors'
{\it Journal of Physics A\/} {\bf 27}, L829--L830.

\item Peres, A. (1993) {\it Quantum Theory: Concepts and Methods\/}
(Dordrecht: Kluwer Academic Publishers).

\item Redhead, M. (1987) {\it Incompleteness, Nonlocality, and
Realism\/} (Oxford: Clarendon Press).

\item Santos, E. (1991) `Does Quantum Mechanics Violate the Bell
Inequalities?' {\it Physical Review Letters} {\bf 66}, 1388--1390.

\item Santos, E. (1992) `Critical Analysis of the Empirical Tests of
Local Hidden-Variable Theories' {\it Physical Review A\/} {\bf 46},
3646--3656.

\item Stapp, H. P. (1972) `The Copenhagen Interpretation' {\it American
Journal of Physics\/} {\bf 40}, 1098--1116.

\item Webster (1974) {\it New World Dictionary of the American
Language\/} (New York: Collins) p.~737.

\end{description}
\end{document}